\documentclass[english,aps,preprint]{revtex4-1}
\usepackage[T1]{fontenc}
\usepackage[latin9]{inputenc}
\usepackage{textcomp}
\usepackage{graphicx}

\makeatletter

\providecommand{\tabularnewline}{\\}

\@ifundefined{textcolor}{}
{%
 \definecolor{BLACK}{gray}{0}
 \definecolor{WHITE}{gray}{1}
 \definecolor{RED}{rgb}{1,0,0}
 \definecolor{GREEN}{rgb}{0,1,0}
 \definecolor{BLUE}{rgb}{0,0,1}
 \definecolor{CYAN}{cmyk}{1,0,0,0}
 \definecolor{MAGENTA}{cmyk}{0,1,0,0}
 \definecolor{YELLOW}{cmyk}{0,0,1,0}
 }

\makeatother

\usepackage{babel}

\begin{document}

\title{Tunability of the reflection and the transmission spectra of two
periodically corrugated metallic plates, obtained by control of the
interactions between plasmonic and photonic modes}

\author{Avner Yanai}

\author{Uriel Levy}

\email{ulevy@cc.huji.ac.il}

\homepage{www.cs.huji.ac.il/~ulevy}

\affiliation{Department of Applied Physics, The Benin School of Engineering and
Computer Science, The Hebrew University of Jerusalem, Jerusalem, 91904,
Israel }
\begin{abstract}
We theoretically study the interactions between plasmonic and photonic
modes within a structure that is composed of two thin corrugated metallic
plates, embedded in air. We show that the interactions depend upon
the symmetry of the interacting modes. This observation is explained
by the phase difference between the Fourier components of the two
gratings. The phase can be controlled by laterally shifting one grating
with respect to the other. Therefore, this relative shift provides
an efficient \textquotedblleft{}knob\textquotedblright{} that allows
to control the interaction between the various modes, resulting in
an efficient modulation of light transmission and reflection in the
proposed structure. Based on this concept we show that the investigated
structure can be used as tunable plasmonic filter.
\end{abstract}
\maketitle

\section{introduction}

The field of surface plasmon polaritons (SPPs) {[}1{]} is rapidly
developing over the last couple of decades. One of the active plasmonic
related research topics is the waveguiding characteristics of multilayered
plasmonic structures {[}2{]}. A basic example for such a structure
is that of a thin metal film sandwiched between two dielectric (insulating)
media (IMI). For a thin enough film, the SPP modes guided by the two
dielectric-metal interfaces are coupled through the metal, thus creating
supermodes that exhibit a dispersion varying with metal thickness.
A variant of the IMI structure that has been studied recently is a
doubly corrugated metallic layer which was analyzed for sinusoidal
{[}3,4{]} and rectangular {[}5,6{]} gratings, with possible applications
for a band-gap plasmonic laser and optical filters. A more complex
multilayered configuration is the double metal plate structure, comprising
of an insulator/metal/insulator/metal/insulator (IMIMI) interface.
The dispersion equations and the waveguiding characteristics of this
configuration have been studied by {[}7-9{]}. This structure was recently
applied for the calculation of the optical forces between the metallic
plates {[}10{]}. In this paper, we study a symmetric one-dimensional
IMIMI structure, of which each of the two metallic layers is periodically
corrugated. We analyze this configuration and show that the relative
shift between the corrugated interfaces controls the interaction between
the modes supported by the structure. Furthermore, we demonstrate
that the control of these interactions enables tunable filtering properties
of both the reflection and the transmission spectra. The computer
simulations used for this study are based on the Rigorous Coupled
Wave Analysis (RCWA) method, also known as the Fourier Modal Method
(FMM). We apply the factorization rules that lead to faster convergence
for TM polarization {[}11-14{]}. The paper is structured as follows.
In Section 2, the modes supported by the structure are described and
classified into groups. In Section 3, it is demonstrated how the interactions
between the plasmonic and the photonic modes form an effective \textquotedbl{}absorption
gap\textquotedbl{}, both under normal and oblique incidence. In Section
4, we show how the shift in absorbance lines, can be utilized to obtain
tunable filtering properties of the reflection and transmission spectra.
Furthermore, we show (to our knowledge for the first time) how the
poor transmission can be enhanced by introducing a large refractive
index contrast between the substrate/superstrate and the air gap separating
between the metallic plates.

\section{Modes of an IMIMI structure}

Fig. 1(a) shows a flat double plate structure, with metal layer thickness
$H_{M}$ and dielectric gap between the plates $H_{A}$. As previously
analyzed {[}7-10{]}, a symmetric double metal plate supports four
plasmonic modes. These modes can be classified as two long range surface
plasmon polariton (LRSPP) modes which can be either symmetric or anti-symmetric
with respect to each other (defined as LRS and LRA) and two short
range surface plasmon polariton (SRSPP) modes which can also be either
symmetric or anti-symmetric with respect to each other (SRS and SRA).
Throughout this paper the plane of symmetry is assumed to be in the
middle of the dielectric gap $H_{A}$ (z=0) and the symmetry is defined
with regard to the magnetic field Hy. The characteristic equation
for the symmetric plasmonic modes in a symmetric IMIMI structure embedded
in a uniform dielectric media is given by: 

\begin {equation}\tanh{(k_{D} H_{A} / 2)} = - \frac 
{\frac{k_{D}k_{M}} {\varepsilon_{D}k_{M}} + (k_{M} / \varepsilon_{M})^{2} \tanh{(k_{M}H_{M})}}
{\frac{k_{D}k_{M}} {\varepsilon_{D}k_{M}} + (k_{D} / \varepsilon_{D})^{2} \tanh{(k_{M}H_{M})}}
\end {equation}where $k_{i}^{2} = k_{X}^{2}-\varepsilon_{i} k_{0}^{2}$ is the decay
constant along the Z direction and i = M,D for the metallic and dielectric
layers respectively. To obtain the anti-symmetric modes, the term
$\tanh{(k_{D} H_{A} / 2)}$ should be replaced with $\coth{(k_{D} H_{A} / 2)}$.
Besides these four modes, the structure also supports Fabry-Perot
modes (FPM) with $k_{X}$=0 that reside between the two metal plates.
Also, guided modes exist within the dielectric gap between the plates.
For a not too thin metallic layer thickness, these modes can be approximately
regarded as the TM metallic slab waveguide. As will be shown, the
FPM and the TM guided modes have an important role when considering
potential applications of the investigated structure. Next we add
a rectangular grating modulation with grating depth of $H_{G}$ (either
outwards or inwards, see Fig 1. (b) and (c)). Thus, the permittivity
function takes the form: $\varepsilon(x)= \sum_{h=-\infty}^{\infty} \varepsilon_{h} \exp{(j2\pi hx/L)}$
where L is the grating period. The Fourier components of the permittivity
function are given by: \begin {equation} 
\varepsilon_{h} = [(\varepsilon_{M} - \varepsilon_{D}) \times \sin (\pi h d / L) / (\pi h)] \times 
\exp (j 2 \pi  h S / L)
\end {equation} Where d/L is the duty cycle of the metallic ridge and S is the relative
shift between the two gratings. Under normally incident illumination,
the allowed k-vectors of the modes take discrete values of the multiples
of the grating vector K (K=2$\pi$/L). 

\begin{figure}
\includegraphics[scale=0.7]{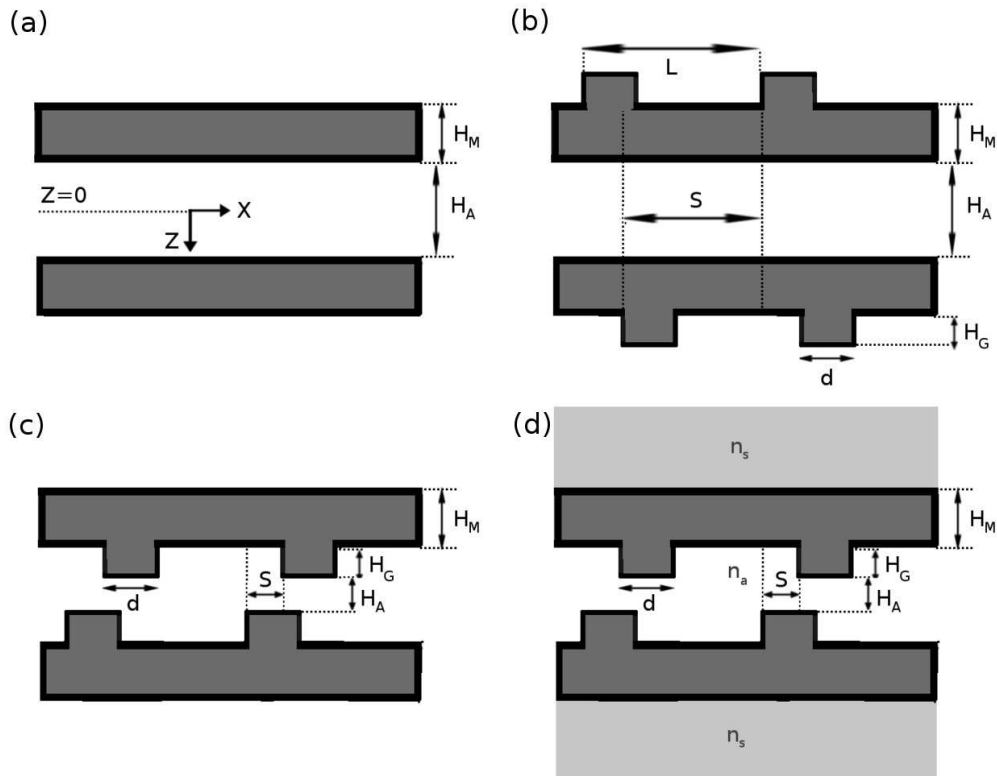}

\caption{Schematic drawing of an IMIMI structure. (a) flat interfaces. (b)
\textquotedbl{}outwards\textquotedbl{} grating modulation. (c) \textquotedbl{}inwards\textquotedbl{}
grating modulation. (d) non-homogeneous dielectric environment with
\textquotedbl{}inwards\textquotedbl{} grating modulation.}

\end{figure}

\section{Mode coupling and interaction}

A 1D periodic structure has a bandgap which resides at the edges of
the Brillouin Zone. It was shown that for a periodic structure with
grating vector 1K, no bandgaps are formed at normal incidence ($k_{X}$=0
), unless the grating has an additional 2K Fourier component {[}3{]}.
As can be observed from Eq. 2, for the specific case of d/L=0.5, the
even Fourier terms vanish and thus no 2K components exist. Therefore,
for such a case no bandgaps should be formed at normal incidence.
This issue and its consequences will be addressed again in Section
4. When considering a double plate structure in which each of the
plates is modulated by a grating (Fig 1(b) and (c)), an additional
\textquotedblleft{}gap\textquotedblright{} mechanism arises. This
\textquotedblleft{}gap\textquotedblright{} is due to mode conversion
as a result of phase matching between two distinct modes. The phase
matching is highly dependent on the mode symmetry as will be shown
immediately. This mode conversion mechanism can be explained by the
Coupled Mode Theory {[}15,16{]} and is well known in photonic structures
{[}17, 18{]}. It was also observed in an adiabatically varying plasmonic
structure {[}19{]}, and by the interaction between a waveguide mode
and a plasmonic mode {[}20{]}.

\subsection{Weak interaction between symmetric and anti-symmetric modes}

To demonstrate the above, we investigated a double plate structure
of the type shown in Fig. 1(b) with the parameters $H_{G}$=80 nm,
$H_{M}$=30 nm, L=1000 nm and d/L=0.75. The materials are assumed
to be air and Ag, i.e. $\varepsilon_{D}$=1 and $\varepsilon_{M}$
is defined by the Drude model $\varepsilon(\omega) = \varepsilon_{\infty} -(\varepsilon_{0}-\varepsilon_{\infty}) \times \omega_{P}^{2}/(\omega^{2} + i \omega \gamma)$
with the following parameters: $\varepsilon_{\infty}$=4.017, $\varepsilon_{0}$=4.896,
$\omega_{P}$=1.419\texttimes{}10$^{16}$ rad/sec, $\gamma$=1.117\texttimes{}10$^{14}$
rad/sec. In Fig. 2, the calculated absorption of this structure under
normally incident TM plane wave illumination is plotted for three
different values of S/L as a function of the incident wavelength and
the separation distance $H_{A}$. The absorption (Ab) is calculated
by the RCWA algorithm using the relation Ab=1\textendash{}T\textendash{}R
where T and R are the total transmission and reflection diffraction
efficiencies respectively. In the absence of absorption, T+R=1. Three
different modes can be observed, designated as A, B and C. Mode A
is the SRS mode. Mode B is the first order FPM and is therefore anti-symmetric
with regard to the magnetic field inside the air gap between the plates.
Mode C is an anti-symmetric guided mode. This mode is not a solution
of the characteristic equations of the non-modulated IMIMI structure
(Eq.1), and can be approximated as the slab $TM_{1}$ mode. The condition
for exciting this mode is L=m$\lambda_{0}$/$n_{eff}$ where m is
an integer and $n_{eff}$ is the effective index of the mode, which
is smaller than one for the wavelengths calculated in Fig. 2. 

\begin{figure}
\includegraphics[scale=0.7]{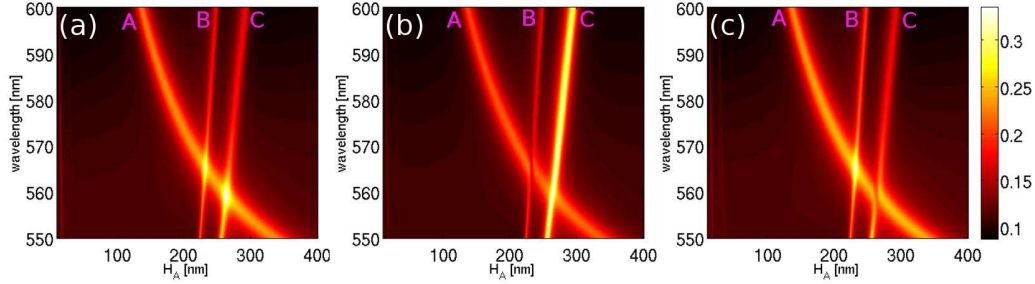}

\caption{Absorption as a function of the incident wavelength and H$_{A}$ for
three different relative shifts between the metal plates for the case
of outwards pointing gratings as shown in Fig. 1(b). (a) S/L=0 (b)
S/L=0.25 (c) S/L=0.5.}

\end{figure}
Thus wavelengths smaller than the period of the grating (1 $\mu$m)
excite this mode. Fig. 3 shows the magnetic field distribution of
the three modes, calculated for a vacuum wavelength of 600 nm. It
can be seen that the SRS mode has an effective wavelength of L/2 (Fig.
3(a)), whereas the FPM (Fig. 3(b)) is invariant along the x-axis,
indicating the absence of a kx component. Fig. 3(c) shows that the
effective wavelength of the $TM_{1}$ slab mode occupies a single
unit cell indicating that the effective index is smaller than 1 (the
unit cell is larger than the vacuum wavelength). To conclude, all
three modes have different $k_{X}$-vectors: for mode A $k_{X}$=2K,
for B $k_{X}$=0K and for C $k_{X}$=1K. In all three insets of Fig.
2 mode A intersects with modes B and C. However, the intersections
result in different interactions in the three considered cases. In
Fig. 2(a) there is no interaction between the different modes and
the absorption at the intersection of the A-B and A-C modes is simply
the summation of the absorption of the two relevant modes. Moreover,
the modes do not alter their characteristics at the intersection region
or in its surrounding. In Fig. 2(b), one can observe an interaction
between the modes, in the form of mode conversion at the A-B intersection.
This results in an anti-crossing between modes A and B. The A-C intersection
is kept unchanged, i.e. no interaction between these two modes is
observed. Fig. 2(c) shows the opposite scenario, where the A-B modes
are not interacting whereas anti-crossing is observed around the A-C
intersection. Let us now describe the mechanism that is affecting
the interaction between the modes. Due to the fact that mode A is
symmetric with $k_{X}$=2K while mode B is anti-symmetric with $k_{X}$=0K,
the phase matching that allows the A\ensuremath{\leftrightarrow}B
transition to occur, involves interaction with the 2K grating component
(more generally, other interactions, e.g. 0K$_{FPM}$+2K$_{SRS}$=3K-1K
might be possible as well, but are weaker than the \textquotedbl{}straightforward\textquotedbl{}
0K$_{FPM}$+2K$_{SRS}$ =2K interaction, and are therefore not considered).
\begin{table}
\begin{tabular}{|c|c|c|c|c|c|}
\hline 
S/L & $\Phi(0K)$ & $\Phi(1K)$ & $\Phi(2K)$ & A\ensuremath{\leftrightarrow}B (2K \ensuremath{\leftrightarrow}0K) & A\ensuremath{\leftrightarrow}C (2K\ensuremath{\leftrightarrow}1K)\tabularnewline
\hline
\hline 
0 & 0  & 0 & 0 & no PM & no PM\tabularnewline
\hline 
0.25 & 0 & $\pi/2$ & $\pi$ & PM & partial PM\tabularnewline
\hline 
0.5 & 0 & $\pi$ & 0 & no PM & PM\tabularnewline
\hline
\end{tabular}

\caption{Phases of the three modes and phase matching between the different
modes for three values of relative grating shift. PM stands for phase
matching}

\end{table}
In Table 1, the values of the phase $\Phi$ (($\Phi$(hK) is the phase
of the hth Fourier component, as calculated from Eq. 2) are summarized
for the three different values of S/L that were considered in Fig.
2. First, we consider the A\ensuremath{\leftrightarrow}B transition.
As modes A and B are of opposite symmetry, no interaction is possible
unless they undergo a relative phase shift of $\pi$ with respect
to the mirror plane z=0. This is similar to the condition for coupling
of symmetric and anti-symmetric modes in photonic grating couplers
(see e.g. {[}18{]}). For S/L=0, this condition can not be satisfied.
In this case the gratings have no phase difference with respect to
each other and therefore the necessary $\pi$ phase shift can not
be provided. However, for S/L=0.25, $\Phi$(2K) undergoes a $\pi$
phase shift. This explains the anti-crossing in Fig. 2(b). For S/L=0.5,
$\Phi$(2K) has again the same phase as for S/L=0. Therefore, no anti-crossing
is seen in Fig. 2(c). For the A\ensuremath{\leftrightarrow}C transition,
the considerations are similar, only now the interaction is provided
by the 1K component. Again, for S/L=0, no interaction is possible,
as the A and C modes have opposite symmetry. As shown in Table 1,
the interacting grating component (i.e. 1K) undergoes a $\pi$ phase
shift for S/L=0.5. Therefore, for these modes we see an anti-crossing
in Fig. 2(c). In Fig. 2(b), only a partial phase match (phase difference
of $\pi$/2) is obtained for the A\ensuremath{\leftrightarrow}C transition,
and no clear anti-crossing can be observed.

\begin{figure}
\includegraphics[scale=0.7]{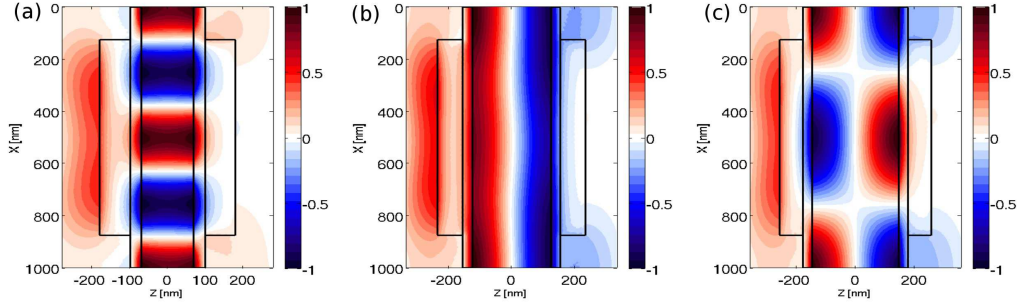}

\caption{Normalized real part of the magnetic field distribution (Hy) calculated
at $\lambda$=600 nm and S/L=0 for: (a) SRS mode (H$_{A}$=137.5 nm).
(b) FPM (H$_{A}$=248.5 nm). (c) TM$_{1}$ mode (H$_{A}$=293 nm).
The square rectangles define the boundaries of the metallic grating
ridges and the metallic plates.}

\end{figure}

\subsection{Strong interaction between modes of the same symmetry}

When the gratings are pointing \textquotedbl{}outwards\textquotedbl{}
(with regard to the dielectric region between the two metal plates)
as in the previously described case, the anti-crossings observed in
Fig. 2 are generally weak. This is because the interactions between
the modes occur mostly in the dielectric region between the plates,
while the gratings are placed in the opposite sides of the metal plates.
Now we consider the case where the gratings are pointing inwards as
shown in Fig.1 (c). The structural parameters are identical to the
previously discussed structure, (subsection 3.1) besides that H$_{G}$=30
nm, with the gratings extending into the dielectric gap between the
plates. The absorption curves of this structure are plotted in Fig.
4(a)-(c), for three different values of S/L. Fig. 4(d) shows schematically
the curves of the original unperturbed modes as they would approximately
exist without inter-modal interaction, superimposed on the S/L=0 case
(also shown in Fig. 4(a)). These unperturbed modes are marked by the
green, blue and white lines. As in the previous simpler case, we have
three modes A, B and C which are the SRSPP modes, FPM, and waveguide
modes (WGM), respectively, where the subscripts denote the symmetry.
In this example we are considering a larger domain both in wavelength
and in separation between the plates. As a result, we can now observe
both symmetries of the three modes. In Fig. 4(d), A$_{A}$ and A$_{S}$
represent the SRA and SRS modes respectively.%
\begin{figure}
\includegraphics[scale=0.7]{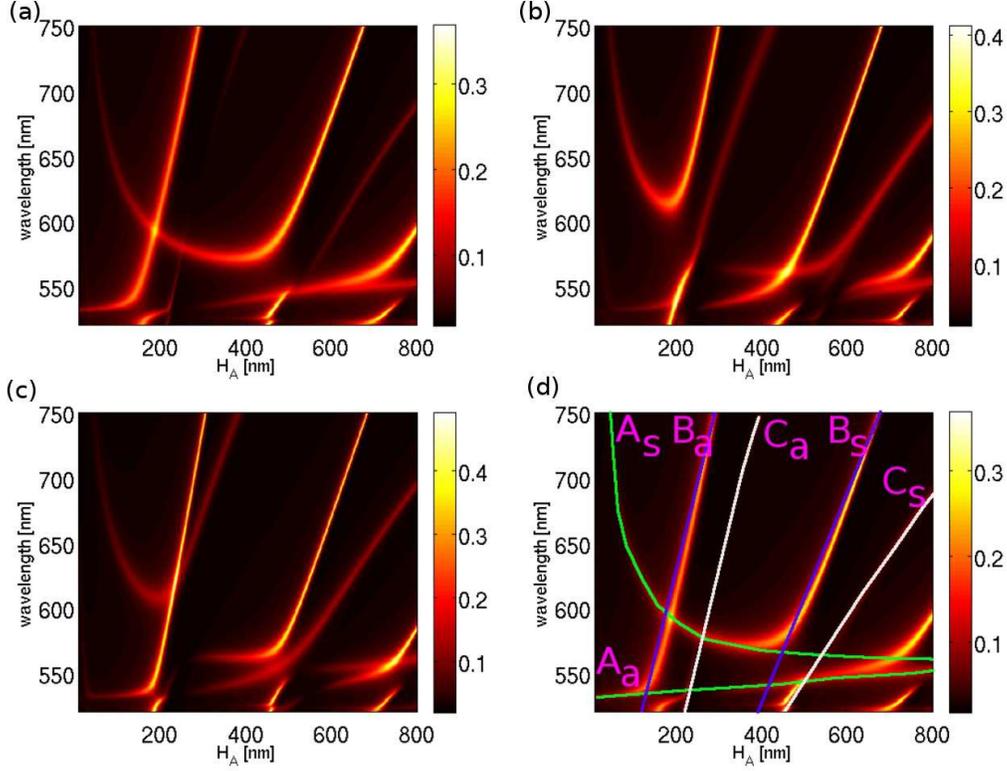}

\caption{Absorption as a function of the incident wavelength and H$_{A}$ for
three different relative shifts between the metal plates for the case
of inwards pointing gratings as shown in Fig.1(c). (a) S/L=0 (b) S/L=0.25
(c) S/L=0.5. (d) Schematic drawing of the supported modes as they
would approximately appear with no inter-modal interaction. The schematic
curves are superimposed on the S/L=0 scenario that is also shown in
Fig. 3(a). The green, blue and white lines represent SRSPP modes,
FPM and WGM respectively (both symmetric and anti-symmetric).}

\end{figure}
 As expected, these two modes can be seen to have the same characteristics
in the limit of H$_{A}$\textrightarrow{}\ensuremath{\infty} as the
two IMI modes have no interaction. B$_{A}$ and B$_{S}$ are the first
and second order FPM respectively, having opposite symmetry. C$_{A}$
and C$_{S}$ are WGM (TM$_{1}$ and TM$_{2}$ respectively). As before,
the K vectors of the SRSPP, FPM and WGM are 2K, 0K and 1K, respectively.
In Fig. 4(a), we can observe the interactions A$_{A}$\ensuremath{\leftrightarrow}B$_{A}$,
A$_{S}$\ensuremath{\leftrightarrow}B$_{S}$ and A$_{A}$\ensuremath{\leftrightarrow}C$_{A}$
as expected, since only interactions between modes of the same symmetry
are allowed. The A$_{S}$\ensuremath{\leftrightarrow}C$_{S}$ interaction
cannot be observed. The absence of this transition may be explained
by the strong A$_{S}$\ensuremath{\leftrightarrow}B$_{S}$ transition,
masking other interactions. In Fig. 4(b) the 2K interactions of opposite
symmetry are allowed. Therefore we see the A$_{S}$\ensuremath{\leftrightarrow}B$_{A}$
and A$_{A}$\ensuremath{\leftrightarrow}B$_{S}$ interactions. We
also identify interactions involving the 1K component between modes
of the same symmetry, because the 1K component is partially matched
(i.e. A$_{A}$\ensuremath{\leftrightarrow}C$_{A}$ and also A$_{S}$\ensuremath{\leftrightarrow}C$_{S}$
which in contrary to Fig. 4(a) is now visible as it is not masked
by the A$_{S}$\ensuremath{\leftrightarrow}B$_{S}$ interaction which
is now forbidden). In Fig. 4(c) we see interactions between modes
of opposite symmetry due to the 1K component (A$_{S}$\ensuremath{\leftrightarrow}C$_{A}$,
A$_{A}$\ensuremath{\leftrightarrow}C$_{S}$) and continue to see
interactions between modes of the same symmetry that are due to the
2K component (A$_{A}$\ensuremath{\leftrightarrow} B$_{A}$ and A$_{S}$\ensuremath{\leftrightarrow}
B$_{S}$).

\subsection{Mode hybridization under oblique incidence }

We now consider the effect of changing the angle of incidence. For
small oblique incident angles (incident TM plane wave is tilted in
the X-Z plane), the above discussed plasmonic modes will have two
possible frequencies which solve the SPP characteristic equation (Eq.
1) for a given $k_{X}$. Thus, two branches of the SPP modes originating
from $k_{X}$=0 appear at a dispersion diagram of the structure. This
is in contrast to the FPM mode which still exhibits a single branch,
because the FPM condition for $k_{Z}$ is uniquely satisfied by an
increase of frequency with the increase of the incident angle. As
a consequence of the existence of two separate SPP branches, distinct
plasmonic modes now intersect at $k_{X}$\ensuremath{\neq}0, and may
interact with each other. %
\begin{figure}
\includegraphics[scale=0.7]{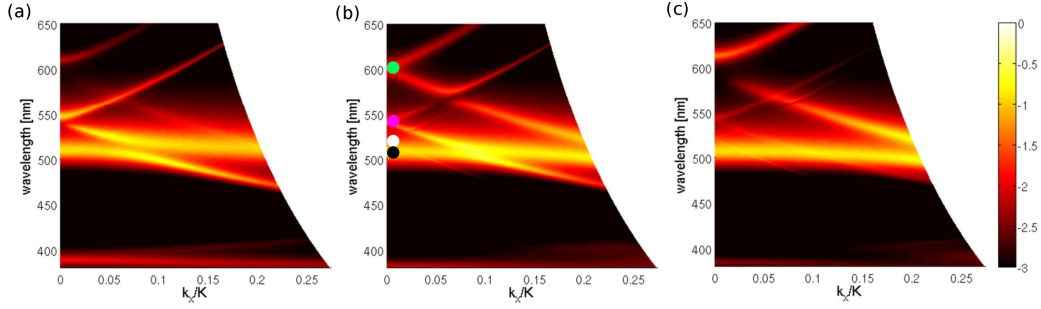}

\caption{Logarithmic scaled plot of the absorption as function of the incident
wavelength and the normalized transverse wavevector k$_{X}$/K for
the following three relative shifts: (a) S/L=0 (b) S/L=0.25. The SRS,
SRA, LRS and FPM are designated at k$_{X}$=0 with green, purple,
white and black dots respectively. (c) S/L=0.5. }

\end{figure}
\begin{figure}
\includegraphics[scale=0.7]{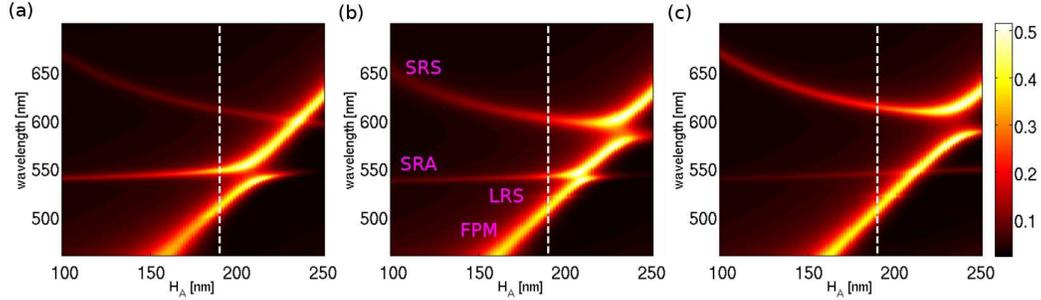}

\caption{The absorption as function of the incident wavelength and the plate
separation under normally incident light for the same structure as
for Fig. 5. The white dashed line corresponds to the H$_{A}$=190
nm at k$_{X}$=0 plotted in Fig. 5. Three relative shifts were considered:
(a) S/L=0 (b) S/L=0.25. (c) S/L=0.5. }

\end{figure}
In Fig. 5, a dispersion diagram is plotted for the \textquotedblleft{}inwards\textquotedblright{}
grating structure with the parameters H$_{A}$=190 nm, H$_{M}$=30
nm, H$_{G}$=10 nm, L=500 nm and d/L=0.75, for three different relative
shifts. Now the grating period is halved compared with the previous
case and thus the SPP modes have a 1K wavevector and there is no coupling
to WGM (the 1K component for L=1000 nm is equivalent to a non existing
0.5K component for L=500 nm). From Figs. 5(a) and 5(c) one can clearly
observe the appearance of a bandgap at $k_{X}$=0 for S/L=0 and S/L=0.5.
This bandgap is due to the 2K grating component, as explained in {[}3,6{]}.
As explained in these references, for S/L=0.25, the bandgaps disappear.
In Fig. 5(b) (where there are no gaps at $k_{X}$=0), we have marked
the excited modes as SRS, SRA, LRS and FPM. The additional LRA mode
is very poorly coupled in this specific configuration and is therefore
not observed (we have observed that for thicker gratings, coupling
to the LRA becomes significant). To help put the current discussion
in the context of the previous section, we also plotted the absorption
of the same structure for normally incident light as a function of
H$_{A}$ (see Fig. 6). The data set at H$_{A}$=190 nm (see vertical
dashed lines in Fig. 6) corresponds to the $k_{X}$=0 case in Fig.
5. Next we discuss the interactions between the modes: 

\textbf{A - Interactions between plasmonic and photonic modes.} The
SRA and FPM interact for S/L=0 around $k_{X}$/K=0.075, since both
modes have the same symmetry. Because the interaction between the
FPM and SPP modes is now obtained through the 1K component, we see
that for S/L=0.5 the FPM and SRS are interacting. Both interactions
can be also observed for S/L=0.25. Yet, the interaction strength is
weaker, because the phase matching is partial. 

\textbf{B - Interactions between the plasmonic modes.} The mechanism
of interactions between plasmonic modes at a single layer for $k_{X}$\ensuremath{\neq}0
was analyzed in {[}21,6{]}, and found to originate from the 2K component.
Therefore, we expect that interaction between distinct plasmonic modes
will occur if the 2K component will provide the required phase shift
to match the symmetry between the plasmonic modes. Thus, interactions
between symmetric and anti symmetric modes should be possible only
for the case of S/L=0.25. Indeed, it is seen that for such a shift
the SRA and SRS modes are interacting whereas for S/L=0 and S/L=0.5
no interaction between these modes can be observed. The interaction
between the LRS and SRS is expected to show an opposite behavior.
This is because both modes have the same symmetry and therefore the
2K component must not provide any phase shift. Indeed, interactions
are observed for S/L=0 and S/L=0.5 but not for S/L=0.25.

\section{Tunable filtering obtained by shifting the relative grating position}

One of the potential applications of the investigated structure is
a tunable filter, where tuning can be obtained by controlling the
relative shift between the two plates. Such a tunable filter has been
previously proposed for multilayered dielectric structures {[}22-24{]}.
Tunable filtering can be obtained for both the transmission and reflection
spectra of the structure. However, while the reflection is significant,
the transmission is low. We will first consider the case of reflection
tunability, and than discuss the modifications needed for obtaining
high transmission that is needed for an efficient tunable transmission
filter. Let us consider a structure with the following parameters:
L=500 nm, H$_{G}$=30 nm, H$_{M}$=30 nm. We choose d/L=0.5 to maximize
the coupling to the SPP modes. As only the 1K component is interacting,
the cases of S/L=0 and S/L=0.5 are the two extreme cases (maximal
phase difference of $\pi$ for the 1K component between these two
cases). %
\begin{figure}
\includegraphics[scale=0.7]{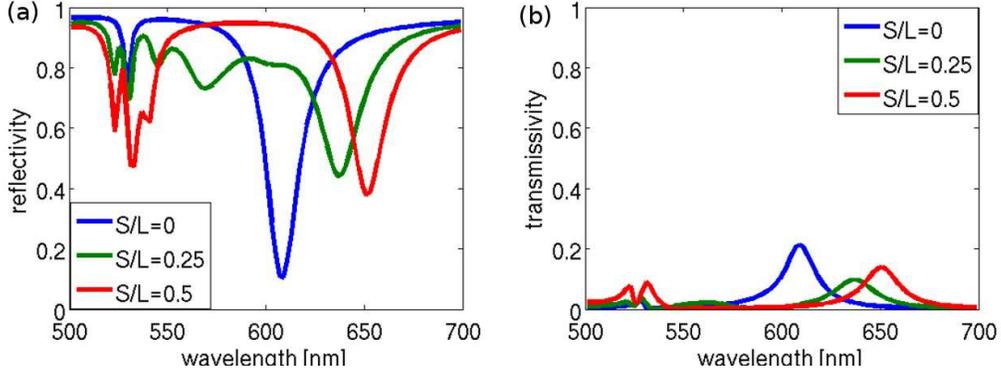}

\caption{(a) Reflectivity and (b) transmissivity as function of the incident
wavelength.}

\end{figure}
In Fig. 7 the reflection and the transmission as a function of wavelength
are plotted for three values of relative shift S/L=0, S/L=0.25, and
S/L=0.5, with separation of H$_{A}$=196 nm. For S/L=0 the first order
FPM and the SRS do not interact, providing low reflection coefficient
of R=0.1 at their crossing point. For S/L=0.5 the modes interact and
the reflection increases to R=0.9 at the same wavelength. In addition,
one can notice that the resonance dip in reflection is shifted in
wavelength. For example, Fig. 7(a) shows a shift of the reflection
dip from 608 nm to $\sim$650 nm. These effects can be used for the
realization of a tunable plasmonic filter. While similar shifts in
the wavelength of resonance are also observed for the transmission
of light through the investigated structure (Fig. 7(b)), the overall
transmission efficiency is seen to be very low. As shown before {[}25,26{]},
the transmission mechanism is via localized SPP (LSP) modes that reside
in the grating ridges, and not through the flat surface SPPs {[}27{]}.
Therefore, in order to enhance the transmission, it is desirable to
confine more energy in the grating ridges, at the expense of lower
energy concentration at the flat surfaces. A possible way to achieve
this is by changing the substrate and superstrate refractive indices
to n$_{S}$=2.6 (e.g. by using Silicon Carbide substrate, see Fig.
1(d) for a schematic of the structure). By keeping the metal layers
thin ($\sim$20 nm) the SPPs on both interfaces remain coupled. The
SPPs tend to be more confined in the lower index dielectric interface,
and to be more radiating at the higher index dielectric. Thereby,
for an \textquotedblleft{}inwards\textquotedblright{} grating configuration,
more energy is confined at the grating ridges that reside near the
lower index material (Fig. 8). In Fig. 8 the transmission and reflection
spectra are plotted for the following configuration: H$_{A}$=25 nm
H$_{M}$=25 nm and H$_{G}$=40 nm. The substrate and the superstrate
have a dielectric index of n$_{S}$=2.6 and the plates are separated
by an air gap (n$_{A}$=1). %
\begin{figure}
\includegraphics[scale=0.87]{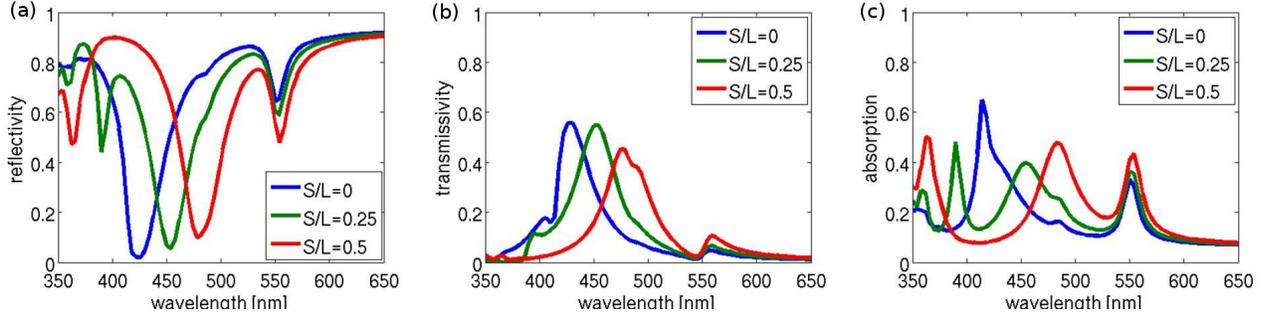}

\caption{(a) reflectivity (b) transmissivity and (c) absorption as function
of the incident wavelength, for the double plate embedded in an inhomogeneous
dielectric index configuration}

\end{figure}
It is observed that the transmission efficiency is greatly enhanced.
The structure still obtains mirror symmetry around Z=0, however the
plasmonic modes can no longer be identified has LRSPP or SRSPP as
these only exist for the cases for which each plate is embedded in
a homogeneous dielectric medium. Still, because of the mirror symmetry,
the modes can be classified as symmetric and anti-symmetric modes
of the overall structure. It can be seen for the absorption spectra
in Fig. 8(c) that an anti-crossing is formed for S/L=0.5. This anti
crossing generates the observed shift in the reflection and transmission
peaks seen in Fig. 8 (a) and (b) respectively.

\section{Conclusions}

We study the plasmonic and photonic modes that are supported by an
IMIMI structure made of thin metallic layers. It is shown that by
adding grating modulation to both metallic layers, the supported modes
can interact. This interaction is explained by the symmetry of the
modes and the relative phase shift provided by the grating Fourier
components. The various interactions are explored both under normally
and oblique incident illumination. Finally, we show that a relative
lateral shift between the two gratings provides tunable filtering
properties. Because of the diversity of the supported modes and their
interactions, this structure seems to be of interest for further research,
and for investigating additional applications, e.g. the selective
excitation of plasmonic modes for plasmonic focusing applications
{[}28,29{]}. 
\begin{acknowledgments}
The authors acknowledge the support of the Israeli Science Foundation,
the Israeli Ministry of Science, and the Peter Brojde Center for Innovative
Engineering and Computer Science.\end{acknowledgments}

\end{document}